\documentclass{iau} 
\usepackage{natbib,aas_macros}
\usepackage{graphicx}

\title[Planet Host Star Magnetism] 
{Magnetism and Activity of Planet Hosting Stars}

\author[Jason T.\ Wright \& Brendan P.\ Miller]   
{Jason T.\ Wright$^1$ and Brendan P.\ Miller$^2$}

\affiliation{$^1$Center for Exoplanets and Habitable Worlds\\and\\[\affilskip]
Department of Astronomy and Astrophysics\\525 Davey Laboratory, The
Pennsylvania State University, University Park, PA 16802 \\ email:
{\tt astrowright@gmail.com}\\
$^2$ Department of Chemistry and Physical Sciences, 
The College of St. Scholastica, Duluth, MN 55811
}
\pubyear{2015}
\volume{320}  
\jname{Solar and Stellar Flares and Their Effects on Planets}
\editors{}
\begin{document}

\maketitle

\begin{abstract}
The magnetic activity levels of planet host stars may differ from that of stars not known to host planets in several ways. Hot Jupiters may induce activity in their hosts through magnetic interactions, or through tidal interactions by affecting their host's rotation or convection. Measurements of photospheric, chromospheric, or coronal activity might then be abnormally high or low compared to control stars that do not host hot Jupiters, or might be modulated at the planet's orbital period. Such detections are complicated by the small amplitude of the expected signal, by the fact that the signals may be transient, and by the difficulty of constructing control samples due to exoplanet detection biases and the uncertainty of field star ages. We review these issues, and discuss avenues for future progress in the field.
\end{abstract}

\firstsection 
\section{Chromospheric Activity as a Confounder In Radial Velocity Searches for Planets}

\subsection{Jitter and RV-activity Correlations}

Studies of stellar magnetic activity of potential planet host stars
are almost as old as radial velocity planet searches.
\citet{Campbell91} reported the discovery of a correlation 
between differential precise radial velocities and magnetic activity
level among active stars.  \citet{Walker95}
showed that the phenomenon was usually not important in their sample
of 21 dwarf stars, but did see strong correlations in 10 years of
precise $\kappa^1$ Ceti radial velocities.  This phenomenon is now
known to be ubiquitous, at a low level (see, for instance, \citet{Lovis11}, \citet{AlphaCenBb}).

Because the correlation between activity level and radial velocity is
imperfect, and in some cases absent despite large variance in both
quantities, active stars have generally been avoided in radial
velocity planet searchers. \citet{Wright04} performed comprehensive analysis of the
chromospheric activity level of stars in the California Planet Survey
at Lick and Keck Observatories, including both overall levels and time series from every
exposure used for radial velocity work, and \citet{Isaacson10}
provided an additional six years of measurements for Keck observations.  The cadence is a few observations per year for most stars, with
higher cadences for bright and multiplanet host stars.  

\citet{Wright04b} used these data in an attempt to
quantify the amount of radial velocity ``jitter'' to be expected from
stars of a given temperature, evolutionary state, and level of
chromospheric activity.  The variance among stars with similar
paramaters in the sample is large, but in general this jitter increases with
activity level, degree of evolution (measured as height above the
$V$-band Main Sequence, $\Delta M_V$), and emission in the core of the
Ca~{\sc ii} H \& K lines.  

A second way that stellar magnetic activity confounds radial velocity
searches for planets is by rotational modulation of starspots. Spots
on the approaching limb of a star can suppress the most blueshifted
component of a star's light, producing a spurious redshift
measurement.  Spots of sufficient size to cause a very large such
shift will also be apparent photometrically as stellar brightness
and color variations with a $\pi/2$ phase shift with respect to the
radial velocities, as in the case of HD
166435 \citep{Queloz166435}.  In this manifestation, the
chromospheric activity enhancement near the spot can sometimes be
also seen with a $-\pi/2$ phase shift from the spurious Doppler
signal.  More complex modeling of this phenomenon, which accounts for
additional effects such as those from plages and the inhibition of convective blueshift, is also possible, as in the 
case, for instance, of FF$^\prime$ \citep{FFprime} and SOAP \citep{SOAP,SOAP2}.

\subsection{Stellar Cycles and Long Period Planets}

\citet{Dravins85} predicted spurious radial velocity
variations due to changes in patterns of surface convection during a
stellar cycle. He recommended tracking precise radial velocities of Solar
lines of different strength, excitation potential, and wavelength to
probe these effects.  \citet{Deming87} reported
precise radial velocities of the Sun in the $\Delta V=2$ transitions
of $^{12}$C$^{16}$O at 2.3$\mu$ in three epochs spanning four years.  They
reported that daily variations were below 3 m/s, but they inferred a
large shift of $-$30 m/s over that span, which they attributed to
the solar cycle.  They inferred that there would therefore be a lower
limit to the mass of planets that could be detected around Sun-like
stars due to this confounding  effect.\footnote{Although, since
  stellar cycles can be monitored, in practice such an effect can be
  detected and removed, and so would not really present an
  insurmountable hurdle to planet detection, even if it were
  pervasive.}
In spite of these predictions and apparent demonstrations that stellar
cycles would be a major problem for radial velocity planet searches,
there have been few manifestations of this problem. The matter was
addressed directly by \citet{Wright08}, who found a correlation among
the radial velocities, activity levels, and brightness of the star HD
154345. They noted that a near spectral twin, $\sigma$ Draconis (HD
185144) had a stronger stellar cycle and higher overall activity
level, and yet serves as a quintessential precise radial velocity
{\it stable} star for
groups around the world.  The lack of an obvious correlation between radial
velocity and activity in $\sigma$ Draconis (and indeed most cycling stars in
the California Planet Survey) and the strength of the RV signal in HD
154345 (over
10 m/s) led those authors to conclude that the correlation was an
inevitable coincidence and that HD 154345 is orbited by
a Jupiter analog (indeed, the first good one discovered). Similarly, \citet{Santos10} found no evidence
for RV-activity correlations due to cycles in a sample of early K dwarfs.

\begin{figure}
\begin{center}
\includegraphics[width=5in]{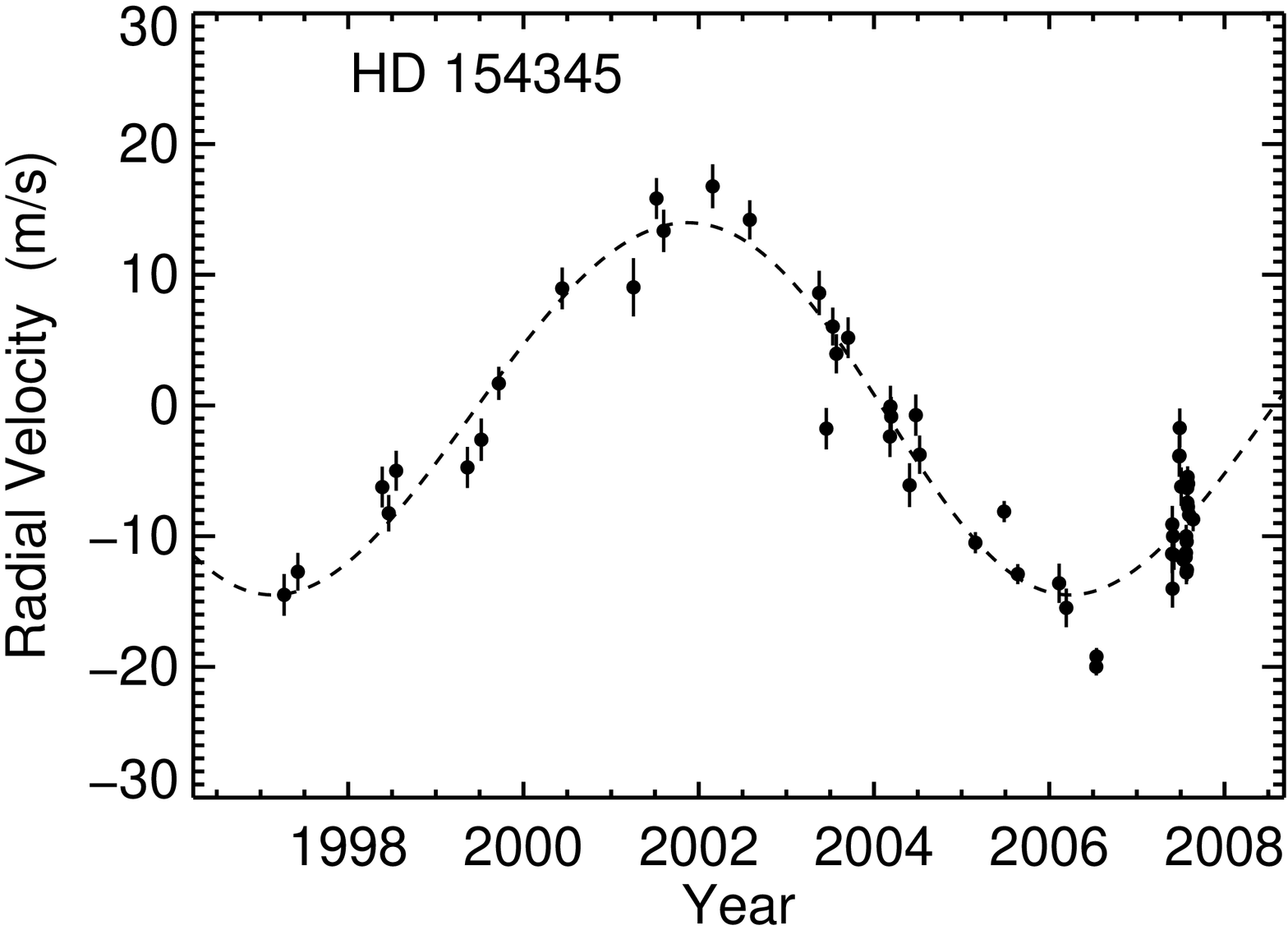}
\includegraphics[width=5in]{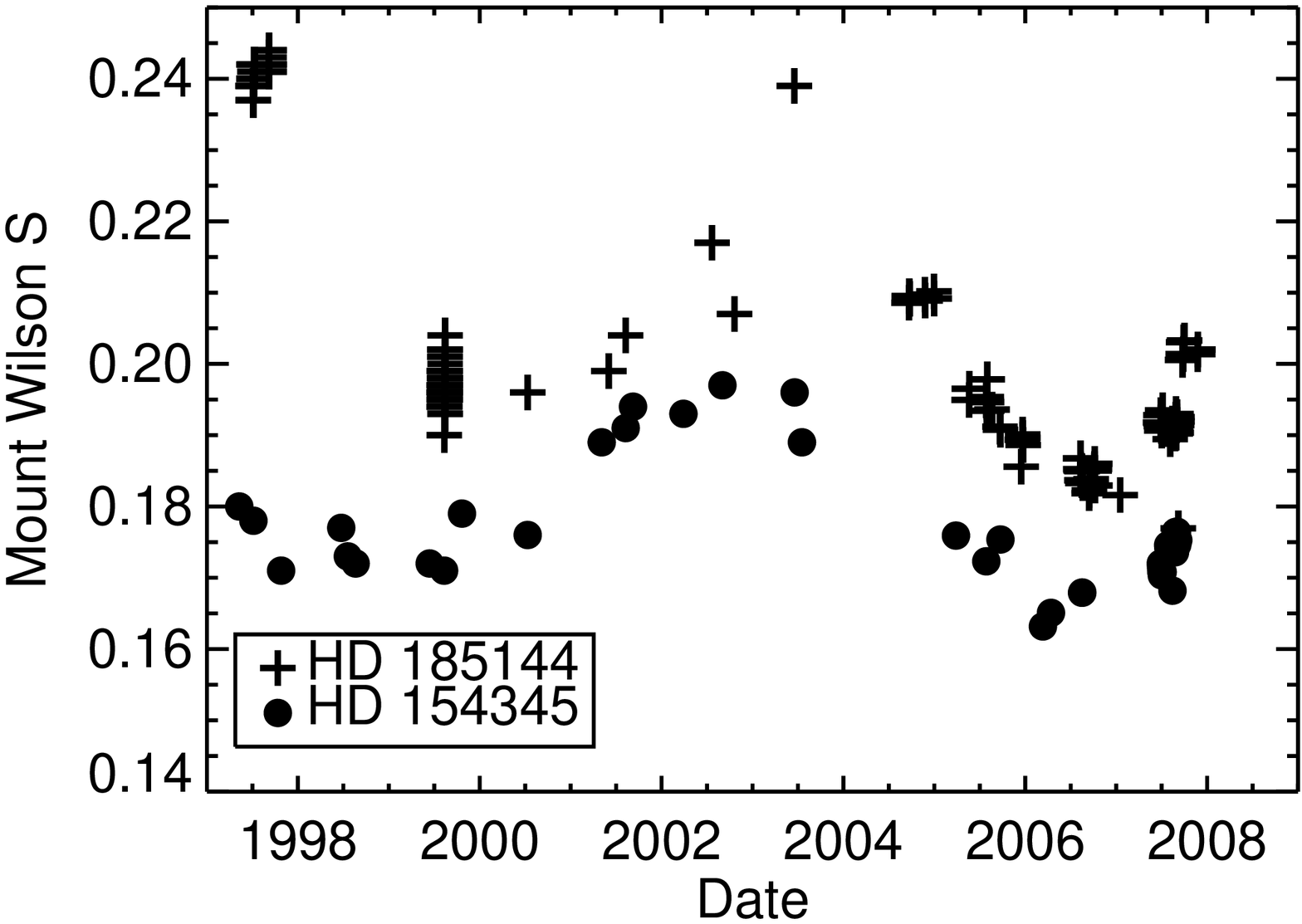}
\caption{Figures 1 (top) \& 2 (bottom) from \citet{Wright08},
  illustrating the strange case of HD 154345, purportedly the first
  good Jupiter analog.  The apparently strong and clean RV signal is
  clearly well correlated with the Mount Wilson $S$ index, a
  measurement of chromospheric activity (circles, bottom).  HD 185144
  (= $\sigma$ Draconis, crosses, bottom) is more active and has a stronger
activity cycle, but is a radial velocity standard star.}
\label{WrightFig}
\end{center}
\end{figure}

Subsequent observations have cast the conclusions of \citet{Wright08}
and \citet{Santos10} into doubt, however,
at least for some stars. \citet{Moutou11} identified two K stars in
their survey with activity-RV correlations with amplitudes $> 10$ m/s
and timescales of years (and confirmed via line bisectors that the
RVs were indeed spurious). \citet{Carolo14} showed a similar
correlation in an active K0 star with an amplitude of $\sim 15$ m/s,
as did \citet{Robertson13} for a K8/M0 star (amplitude $\sim 10$
m/s). Similarly, \citet{AlphaCenBb} showed a 
signal with a $\sim 5$ m/s amplitude in the K1 star $\alpha$ Cen B, and \citet{daSilva12}
found a similar low-level ($<5$ m/s) correlation in a sample of M
dwarfs. Indeed, subsequent observations of HD 154345 have made the
coincidence in that system even stronger, casting doubt on the reality
of the planet.

But the broader point made by \citet{Wright08} and \cite{Santos10}
(and, implicitly, by \citet{daSilva12}) still stands: {\it most}
stars do not show {\it large} RV variations due to activity cycles,
though many may at a low level below that induced by Jupiter-mass
planets. The question of why a loud minority of stars do show strong
correlations that can be misinterpreted as giant planets remains open. 
 
The best comprehensive analysis to date of the effect is the
manuscript of \citet{Lovis11}, who found that RV-activity correlations
due to stellar cycles are more common around hotter stars and almost
absent around cooler stars (a finding contradicted by many of the
citations above), and found {\it average} RV-activity correlation
coefficients as a function of effective temperature and metallicity
(although, as we have seen, there must be a large amount of scatter
about this relation).

\section{Star-Planet Interactions}

\subsection{Modes of Proposed Star-Planet Interactions}
Might the physical connection between activity and RV variation go the other way?
\citet{Jose65} suggested, motivated by the similarity of the Sun's
  $\sim 11$ year activity cycle to the combined orbital action
  of Jupiter and Saturn, that the solar cycle is in some way {\it caused}
  by its planets; \citet{Wilson08} developed this idea further, and
\citet{PerrymanBary} argued that this hypothesis
  could be tested in exoplanetary systems. While the idea has an certain appealing
  counter-intuitiveness to it, \citet{Shirly06} has shown that there is
  really no physical mechanism that could plausibly produce such
  coupling,\footnote{Specifically, \citet{Jose65} and \citet{Wilson08} 
    suggest that the Sun's global field strength is dictated by its
    physical position with respect to the Solar System barycenter.  But by the Equivalence 
    Principle, the gravitational pull of the planets cannot affect the
    Sun's internal motions are any more than an astronaut in a small
    ship can use accelerometers to determine whether they are in 
  orbit or deep space. Any viable mechanism must therefore invoke either 
  gravitational {\it tides} (which are minuscule here) or purely
  electromagnetic effects.} (although \citet{Abreu12} would beg to differ).

Close-in planets, however, could very plausibly interact magnetically
and tidally with their host stars in detectable ways.  Indeed, an
exciting consequence of the stellar magnetic field is how 
it might interact with and so reveal the strength of close-in planetary magnetic
fields.  

\citet{Cuntz00} gave initial estimates of the interaction strengths of
magnetic interactions of stars and their planets, and
\citet{Rubenstein00} suggested the effects of close-in giant planets
might be so large as to be responsible for superflares seen in old FGK
stars. Much work has been done to refine these early
estimates. For instance, \citet{Cohen09} performed time-dependent MHD simulations,
and found that there should be detectable enhancements in X-ray emission and
chromospheric activity \citep[see also][]{Lanza08,Lanza09,Saur13,Cohen11}.

These effects fall into two broad observational classes.  The first
is that a close-in planet might ``tickle'' the star's field, causing orbital energy to be dissipated at the
footprint of the field lines in the chromosphere, resulting in a
chromospheric ``hot spot,'' observed to be modulated on an orbital timescale.
The second is that, through a variety of mechanisms, a close-in
planet might alter the overall level of chromospheric activity.

\subsection{Orbitally Modulated Activity}

The first of these effects can be tested in a few nights' observing
time via a time series of strengths of chromospheric emission lines,
such as the Ca {\sc ii} H \& K lines \citep{Saar01}. \citet{Shkolnik03} claimed the first
detection of this effect as a modulated line strength in the star HD
179949 due to its hot Jupiter companion; the effect was apparently
persistent over three observing runs in 2001 and 2002 and the authors
later claimed to see similar effects in other stars. 

Variation in chromospheric line strengths are expected as
magnetically active regions rotate into and out of view on the star,
but \citet{Shkolnik03} found that they phased better with the orbital
period of the planet than the rotational period of the star (although
\citet{Miller12} showed that with poor phase coverage such
distinctions can be difficult or impossible to make).

\begin{figure}
\begin{center}
\includegraphics[width=4.5in]{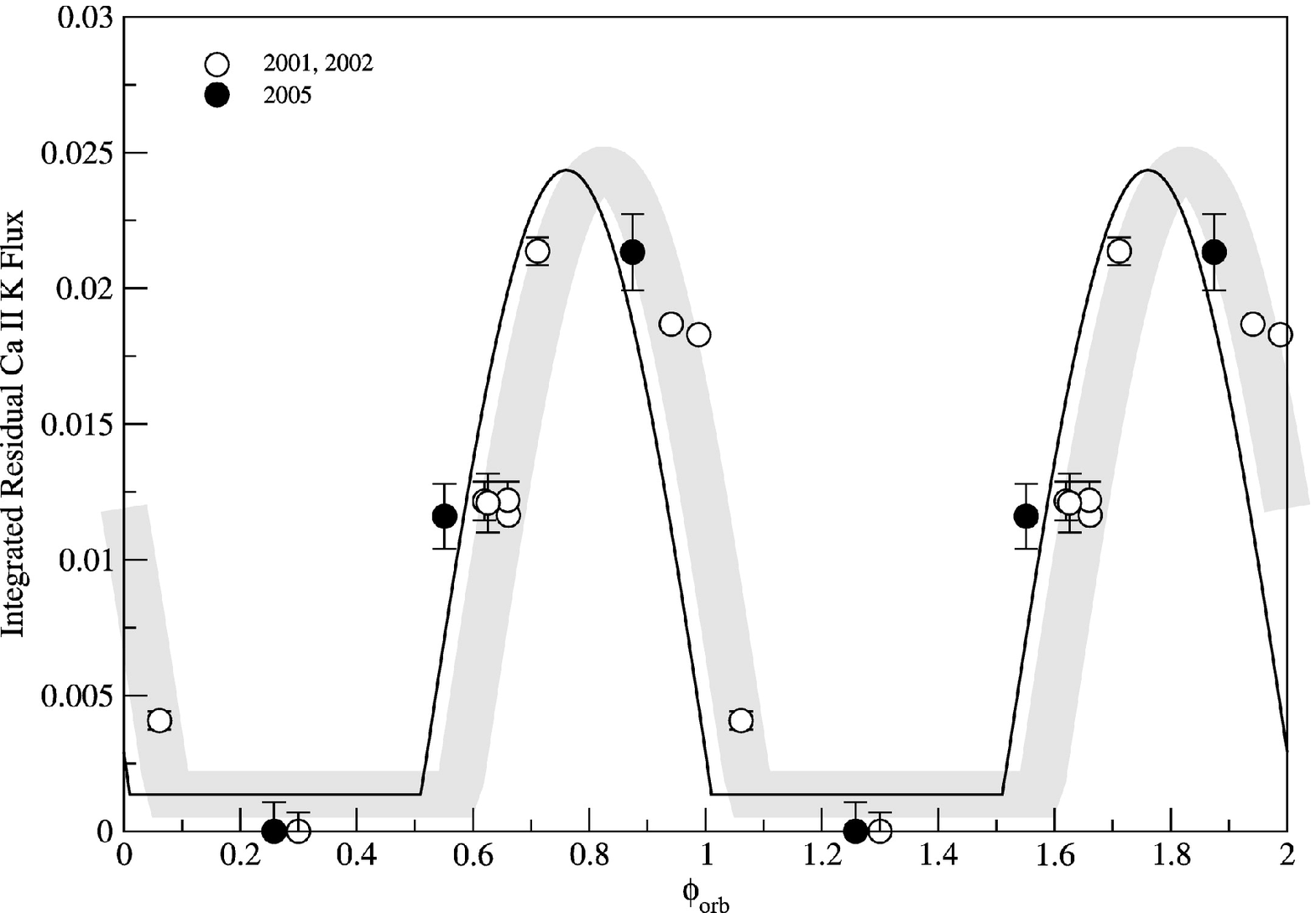}
\includegraphics[width=4.5in]{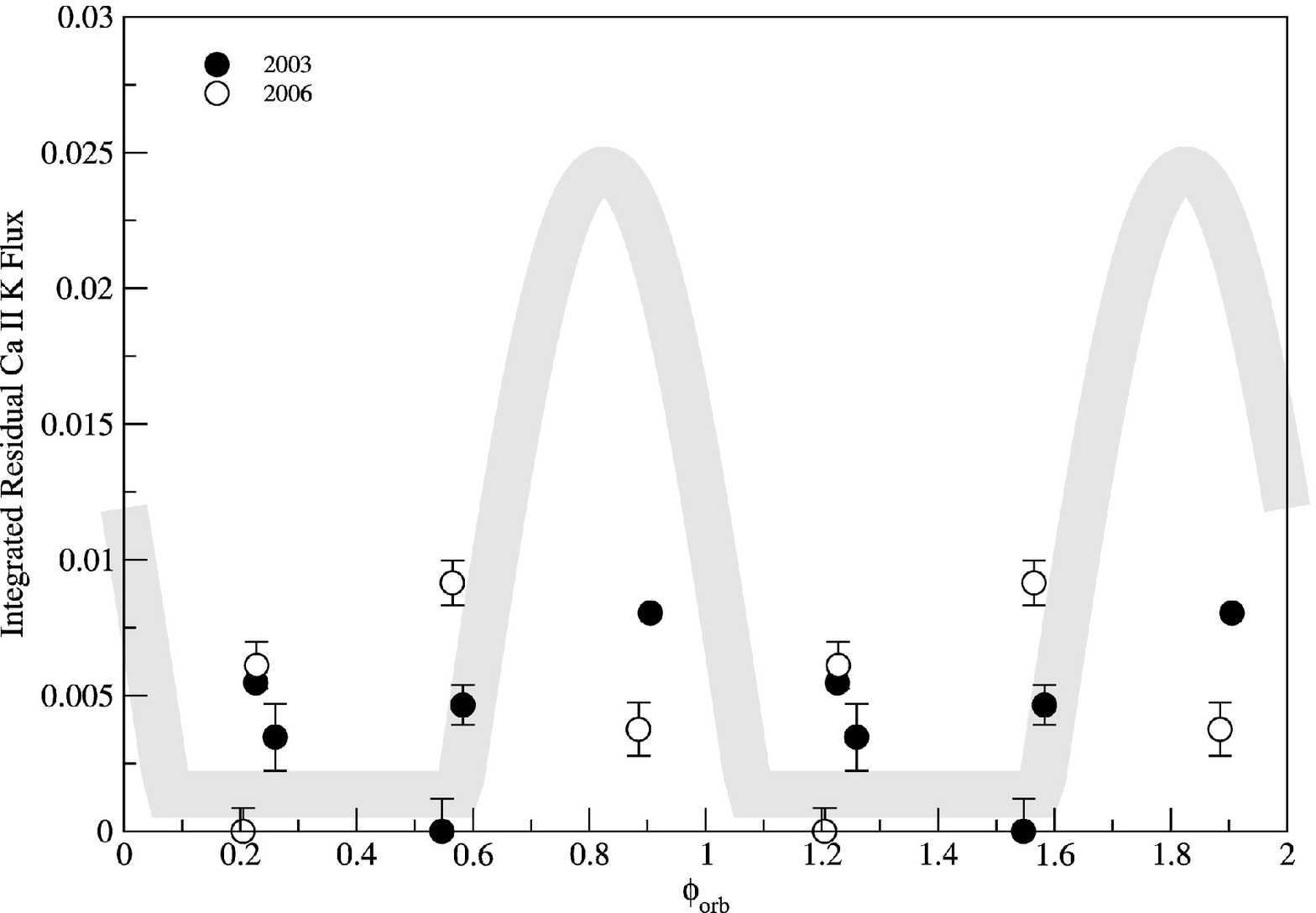}
\caption{Figures 6 (top) \& 7 (bottom) from \citet{Shkolnik08} showing the ``on-off''
  nature of star-planet interactions in HD 179949.  Note that each
  data point appears twice in these plots because phase runs from zero
  to 2 to emphasize the purportedly periodic nature of the signal.  On the left, the ``on'' epochs, which have a slightly different best-fit phase in the
2001--2 epoch (thick gray line) than in the 2005 epoch (black line).
On the right, the ``off'' epochs where no effect is seen.}
 \label{Shkolnik}
\end{center}
\end{figure}

\citet{Shkolnik05} appeared to confirm the result, and to find a
similar effect in the $\upsilon$ And system. \citet{Shkolnik08} found
the detections of these signals difficult to reproduce, leading them
to speculate that the effect has an ``on/off nature'' (see
Figure~\ref{Shkolnik}).  Indeed, \citet{Poppenhaeger11b} found no
variability in the $\upsilon$ And system at the orbital period of its
close-in planet in Ca {\sc ii} H \& K or in X-rays, nor did
\citet{Scandariato2013} in their observing campaign on HD~179949. The
difficulty in confirming this mode of interaction since its putative
discovery casts doubt on the interpretation that the variations were
necessarily due to planetary perturbations.

\subsection{Enhancement of Overall Activity}

Detection of an overall enhancement (or decrease) in activity levels
is more difficult because in general one must know the ``correct''
level for the star, which might have a cycle. Most work on this topic
has thus been done statistically, to determine if stars with hot
Jupiters appear unusually active. Such work requires careful
construction of a control sample, since there are many observational selection
effects in searches for hot Jupiters that can confound such an
analysis \citep{Poppenhaeger11}.

\citet{Kashyap08} performed an analysis of X-ray emission from a
sample of planet-host stars, searching for correlations with orbital
distance (since the strength of the effect should decrease with
increasing orbital distance).  They found that stars with close-in
planets are more X-ray active by a factor of 4 than distant planets,
but this conclusion is subject to several difficulties, including
non-planets orbiting very active stars (e.g.\ they included the brown
dwarf candidate Cha H$\alpha$ 8 B, whose host star is active primarily
because it is 3 Myr old \citep{Joergens10}) and observational biases
(transit surveys sensitive to close-in planets can sample more active
stars, while long-period planets are preferentially discovered around
inactive stars favored by radial-velocity methods). They attempt to
correct for sample bias and estimate a planet-induced activity
enhancement by a factor of 2. However, \citet{Poppenhaeger10}
performed a more careful analysis with a more sensitive set of X-ray
data and found no evidence for elevated activity among hosts of hot
Jupiters.

\citet{Scharf10} tentatively identified a relation between stellar
$L_X$ and planet mass for close-in planets (that is not present for
more distant planets) in {\it ROSAT\/} All-Sky Survey
data. \citet{Poppenhaeger11c} showed that this correlation did not
survive the addition of more sensitive {\it XMM-Newton\/} data, and
that the correlations that had been seen in the past were likely due
to selection effects in the planet-search process.

\begin{figure}
\begin{center}
\includegraphics[width=5in]{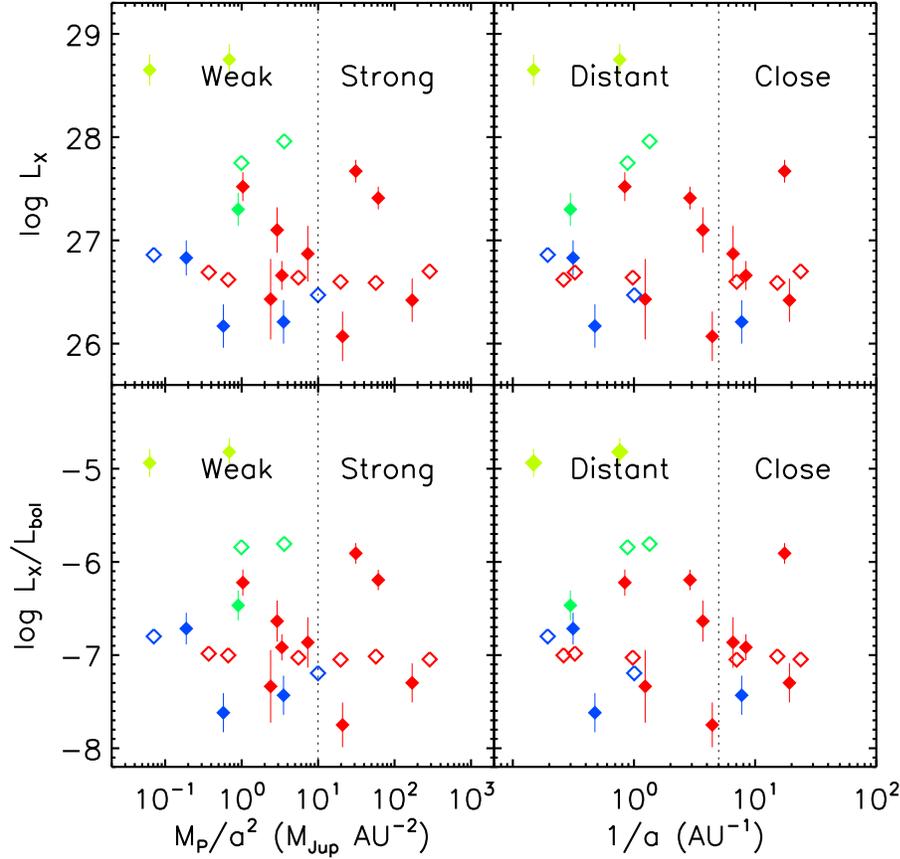}
\caption{Figure 4 of \citet{Miller15}, showing the distribution of
  coronal activity vs.\ plausible scalings for planet-star interaction
  strength for solar analogs in their sample.  Open symbols represent
  upper limits.  {\it Chandra} data are
  red, {\it XMM-Newton} data are blue, and {\it ROSAT} data are
  green.  \citet{Miller15} found no significant increase in X-ray
  emission among the stars most likely to show elevated activity due
  to star-planet interactions.  Light green points were excluded from
  their analysis as being 
  ``atypically active.''}
\label{MillerFig1}
\end{center}
\end{figure}

\begin{figure}
\begin{center}
\includegraphics[width=5in]{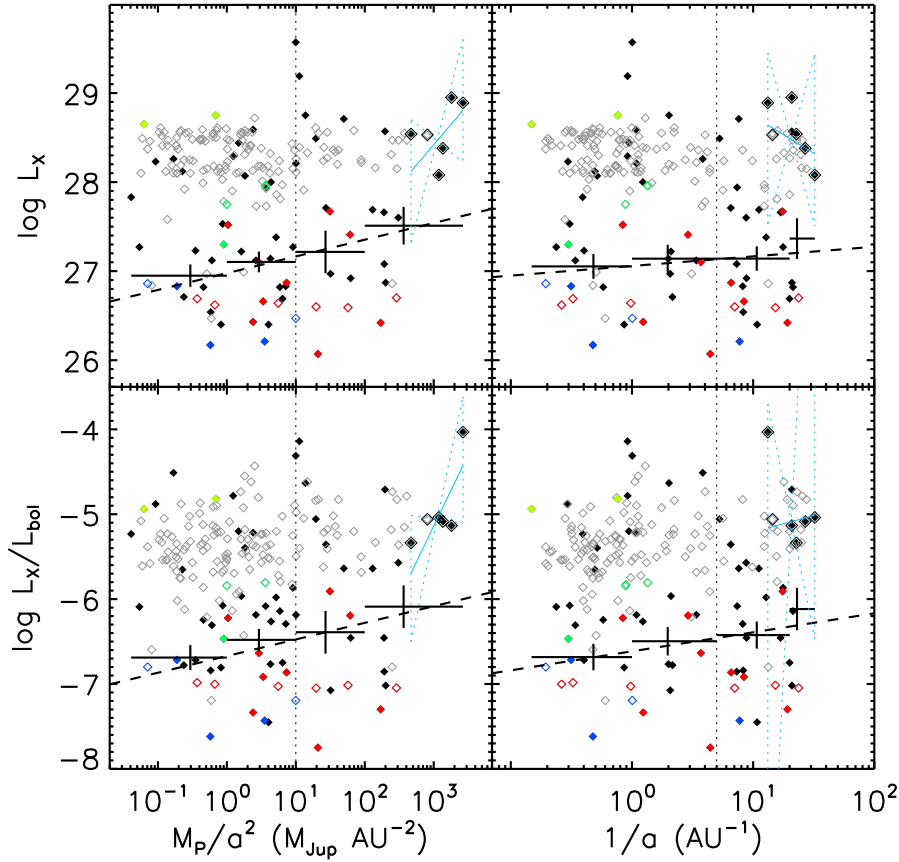}
\caption{Figure 7 of \citet{Miller15}, showing the distribution of
  coronal activity vs.\ plausible scalings for planet-star interaction
  strength for their entire sample, with the same color scheme as
  Figure~\ref{MillerFig1} for the Solar analogs, and other stars in
  black.  The double diamond symbols (which have a fit shown in cyan)
  are ``extreme'' systems which drive the weak correlation shown by
  the black dashed line.}
\label{MillerFig2}
\end{center}
\end{figure}

But more carefully constructed samples have hinted at a more subtle role
for hot Jupiters to play.  \citet{Poppenhaeger14} examined
coronal X-rays from the components of wide binary
stars in which one star hosts a hot Jupiter, using its companion as a
negative control, and found that the Jupiter hosts do seem to be more
active than their coeval sibling. \citet{Miller15} performed a
comprehensive and careful assessment of 
X-ray data of exoplanet hosts, seeking correlations in coronal X-ray
strength with plausible metrics of star-planet magnetic interaction
($M_p/a^2$, $1/a$). They found no significant elevation in a sample of
solar analogs (Figure~\ref{MillerFig1}), but did find a significant correlation driven by
a few very massive, very close-in exoplanets orbiting X-ray bright
stars (Figure~\ref{MillerFig2}). These stars' activity levels were consistent with their
chromospheric emission and rotation rates, and so did not appear
to be elevated because of magnetic interactions with the planet.

\subsection{An Indirect Mode For Magnetic Activity Enhancement Via
  Star-Planet Interactions}

Both \citet{Poppenhaeger14} and \citet{Miller15} argued that the
source of the elevated activity levels is consistent with the stars
having been spun up by their massive, close-in planets. In this
scenario, the normal spin-down of stars with age is slowed or halted
by tidal interactions with their planets, and so their dynamos remain
strong late into the stars' lives.  Such action is instructive for
determining the factors contributing to tidal coupling between stars
and planets, but dampens hopes for using stellar magnetic fields to
probe exoplanetary magnetic fields.

\section{WASP-18 as Case Study}

\subsection{A Search in the Strongest Candidate for SPI Comes Up Empty}

WASP-18 \citep{WASP-18, WASP-18b} is an ultra-short period transiting hot
Jupiter with $M_p =10$ M$_{\rm Jup}$ orbiting an F6 star.  The
brightness of the host star ($V$=9.3) and the mass and orbital
distance of the planet make the WASP-18 system one of the best
candidates for observations of star planet interactions.

Strangely, however, WASP-18 is unusually {\it inactive}, both
chromospherically \citep{Knutson10, Fossati14} and
coronally \citep{Miller12,Pillitteri14}.  It has $\log R^\prime_{\rm HK} = -5.15$,
which is extraordinarily low for a main sequence star, and is more
typical of subgiants \citep{Wright04b}, despite the star having a
rotation period (5.6 days) ordinary for its spectral type and age.

This low level of activity
frustrated the attempts of \citet{Miller12} and \citet{Pillitteri14} to monitor the system for
orbitally modulated activity in X-rays; it was not detected in 50 ks
of {\it Swift} time or in 87 ks of {\it Chandra} time. If this system
does not show enhanced activity due to star-planet interactions,
other, less favorable
systems might not be expected to show anything detectable.

\subsection{A Different Manifestation of Planetary Influence?}

The solution to the puzzle of the low activity levels of WASP-18 may be the
extraordinarily large and close-in planet orbiting it.  One
possibility, mentioned by \citet{Miller12}, is that planetary tides
have in some way disrupted the dynamo action of the star (see also
Wolk, Pillitteri, and Poppenhaeger, in these proceedings).  

A second suggestion, by \citet{Fossati13}, is that the chromospheric
measurements of WASP-18 are contaminated by absorption from
intervening gas escaping from the planet (\citet{Fossati14} later
ruled out ISM absorption).  This mechanism has been suggested by
\citet{Lanza14} and \citet{Fossati15} to explain the correlation,
first identified by \citet{Hartman10}, that low surface gravity hot
Jupiters tend to orbit stars that appear less active. The very high
surface gravity of WASP-18 would seem to cut against such a
possibility in this instance, but the concern is apparently warranted
for WASP-12 \citep{Fossati13}, and presumably other systems as wall.
At any rate, \citet{Pillitteri14} ruled this mechanism out for
WASP-18, since the star is also unusually faint in hard X-rays, which
should be relatively unattenuated by a gas cloud.

\citet{Pillitteri14} note that the star has significant lithium, more
than expected even for a Hyades-aged star. \citet{Pillitteri14}
combine the mystery of the star's high lithium abundance and low
activity with a tidal strength calculation, and infer that WASP-18
{\it b} has indeed disrupted its host star's dynamo via tidal
suppression of convection in the shallow convective zone --- an effect
that also served to frustrate lithium burning at the base of the
convective zone.  This hypothesis would gain support if the
combination of unusually low activity and unusually high lithium
abundance is common in similar stars with very close, massive
companions (i.e.\ stars, brown dwarfs, or very massive planets).  Such
an observation would cut against the general trend of late F stars
hosting hot Jupiters to have {\it lower} lithium abundances than stars
without hot Jupiters \citep{DelgadoMena15}.


\end{document}